\documentstyle[prd,preprint,tighten,aps,eqsecnum,floats,%
amssymb,amsfonts,newlfont,graphicx]{revtex}
\begin{document}
\draft
\preprint{
\begin{tabular}{r}
DFTT 48/99\\
hep-ph/9909465
\end{tabular}
}
\title{Experimental Constraints on Four-Neutrino Mixing}
\author{Carlo Giunti}
\address{INFN, Sezione di Torino, and Dipartimento di Fisica Teorica,
Universit\`a di Torino,\\
Via P. Giuria 1, I--10125 Torino, Italy}
\maketitle
\begin{abstract}
It is shown that at least four massive neutrinos are
needed in order to accommodate the evidences in favor of neutrino oscillations
found in solar and atmospheric neutrino experiments
and in the LSND experiment.
Among all four-neutrino schemes,
only two  are compatible with the results
of all neutrino oscillation experiments.
These two schemes have a mass spectrum composed of
two pairs of neutrinos with close masses 
separated by the ``LSND gap'' of the order of 1 eV.
\end{abstract}

\pacs{Talk presented at
\textit{Neutrino Mixing},
Meeting in Honour of Samoil Bilenky's 70$^{\mathrm{th}}$ Birthday,
Torino, 25--27 March 1999.}

\section{Introduction}
\label{Introduction}

Neutrino oscillations have been proposed by B. Pontecorvo
\cite{Pontecorvo-57-58}
more than forty years ago
from an analogy with $K^0 \leftrightarrows \bar{K}^0$
oscillations.
In 1967 B. Pontecorvo
predicted the possibility that the flux of solar $\nu_e$'s
could be suppressed because of neutrino oscillations
\cite{Pontecorvo-67}.
About one year later
\cite{Davis-68}
R. Davis and collaborators
reported the first measurement of
the Homestake
$^{37}$Cl$(\nu_e,e^-)$$^{37}$Ar
radiochemical detector,
which gave an upper limit for the flux of solar $\nu_e$'s on the Earth
significantly smaller than the prediction
of the existing Standard Solar Model
\cite{Bahcall-68}.
This was the first experimental indication in favor of neutrino oscillations.

It is interesting to notice that in the 1967 paper
\cite{Pontecorvo-67}
B. Pontecorvo introduced the concept of ``sterile'' neutrinos.
He considered the possibility of oscillations
of left-handed neutrinos created in weak interaction processes into
left-handed ``antineutrino'' states,
quanta of the right-handed component of the neutrino field
that does not participate in weak interactions.
He called these states ``sterile'',
opposed to the usual ``active'' neutrinos.
He also pointed out that in this case
the mass eigenstates are Majorana particles.
As we will see later,
the present experimental data
indicate the existence of at least one sterile neutrino.
 
In 1969
V. Gribov and B. Pontecorvo
wrote a famous paper in which they formulated the theory of
$\nu_e\to\nu_\mu$ oscillations.
In 1976 S.M. Bilenky and B. Pontecorvo
\cite{Bilenky-Pontecorvo-76a}
introduced the general scheme
with Dirac and Majorana neutrino mass terms,
lying the foundations
of the theory of neutrino mixing.
The early studies of neutrino mixing
and neutrino oscillations
are beautifully summarized in the 1978 review of
S.M. Bilenky and B. Pontecorvo \cite{Bilenky-Pontecorvo-78}.

Today neutrino oscillations
are subject to intensive experimental and theoretical research
\cite{reviews,BGG-98-review}.
This beautiful quantum mechanical phenomenon
provides information on the masses and mixing of neutrinos
and
is considered to be one of the best ways to explore the physics beyond the
Standard Model.
Indeed, the smallness of neutrino masses
may be due to the existence of a very large energy scale
at which the conservation of lepton number is violated
and small Majorana neutrino masses are generated through the see-saw mechanism
\cite{see-saw}
or through non-renormalizable interaction terms in the
effective Lagrangian of the Standard Model
\cite{effective}.

The best evidence in favor of the existence of neutrino oscillations
has been recently provided by the measurement
in the Super-Kamiokande experiment \cite{SK-atm}
of an up--down asymmetry of high-energy
$\mu$-like events generated by atmospheric neutrinos:
\begin{equation}
\mathcal{A}_\mu
\equiv
(D_\mu-U_\mu)/(D_\mu+U_\mu)
=
0.311 \pm 0.043 \pm 0.01
\,.
\label{Amu}
\end{equation}
Here $D_\mu$ and $U_\mu$ are,
respectively,
the number of downward-going and upward-going events,
corresponding to the zenith angle intervals
$0.2 < \cos\theta < 1$
and
$-1 < \cos\theta < -0.2$.
Since the fluxes of high-energy downward-going and upward-going
atmospheric neutrinos are predicted to be equal with high accuracy
on the basis of geometrical arguments
\cite{Lipari-99},
the Super-Kamiokande evidence in favor of neutrino
oscillations is model-independent.
It provides a confirmation of the indications
in favor of oscillations of atmospheric neutrinos
found through the measurement
of the ratio of $\mu$-like and $e$-like events
(Kamiokande,
IMB,
Soudan 2,
Super-Kamiokande)
\cite{atm-exp-contained,SK-atm}
and through the measurement
of upward-going muons produced by neutrino interactions
in the rock below the detector
(Super-Kamiokande,
MACRO)
\cite{atm-exp-upmu}.
Large
$\nu_\mu\leftrightarrows\nu_e$
oscillations of atmospheric neutrinos
are excluded
by the absence of a
up--down asymmetry of high-energy
$e$-like events generated by atmospheric neutrinos and detected in
the Super-Kamiokande experiment
($
\mathcal{A}_e
=
0.036 \pm 0.067 \pm 0.02
$)
\cite{SK-atm}
and by the negative result
of the CHOOZ long-baseline $\bar\nu_e$ disappearance experiment
\cite{CHOOZ-99}.
Therefore,
the atmospheric neutrino anomaly consists in the disappearance
of muon neutrinos and can be explained by
$\nu_\mu\to\nu_\tau$
and/or
$\nu_\mu\to\nu_s$
oscillations
(here $\nu_s$ is a sterile neutrino
that does not take part in weak interactions).

Other indications in favor of neutrino oscillations have been obtained in
solar neutrino experiments
(Homestake,
Kamiokande,
GALLEX,
SAGE,
Super-Kamiokande)
\cite{sun-exp}
and in the LSND experiment \cite{LSND}.

The flux of electron neutrinos measured in
all five solar neutrino experiments
is substantially smaller than the one predicted
by the Standard Solar Model \cite{BP98}
and a comparison of the data of different experiments
indicate an energy dependence of the solar $\nu_e$ suppression,
which represents a rather convincing evidence
in favor of neutrino oscillations
\cite{BGG-98-review}.
The disappearance of solar
electron neutrinos
can be explained by
$\nu_e\to\nu_\mu$
and/or
$\nu_e\to\nu_\tau$
and/or
$\nu_e\to\nu_s$
oscillations
\cite{sun-analysis}.

The accelerator LSND experiment
is the only one that claims the observation
of neutrino oscillations in specific appearance channels:
$\bar\nu_\mu\to\bar\nu_e$ and $\nu_\mu\to\nu_e$.
Since
the appearance of neutrinos with a different flavor
represents
the true essence of neutrino oscillations,
the LSND evidence is extremely interesting
and its confirmation (or disproof)
by other experiments
should receive high priority in future research.
Four such experiments have been proposed and are under study:
BooNE at Fermilab,
I-216 at CERN,
ORLaND at Oak Ridge
and
NESS at the European Spallation Source \cite{LSND-check}.
Among these proposals only BooNE is approved and will start in 2001.

Neutrino oscillations occur
if neutrinos are massive and mixed particles
\cite{Bilenky-Pontecorvo-78,reviews,BGG-98-review},
\textit{i.e.} if
the left-handed components
$\nu_{{\alpha}L}$
of the flavor neutrino fields
are superpositions of
the left-handed components
$\nu_{kL}$
($k=1,\ldots,N$)
of neutrino fields with definite mass
$m_k$:
\begin{equation}
\nu_{{\alpha}L}
=
\sum_{k=1}^{N}
U_{{\alpha}k}
\,
\nu_{kL}
\,,
\label{mixing}
\end{equation}
where $U$
is a $N{\times}N$ unitary mixing matrix.
From the measurement of the invisible decay width of the $Z$-boson
it is known that the number of light active
neutrino flavors is three
\cite{PDG98},
corresponding to $\nu_e$, $\nu_\mu$ and $\nu_\tau$.
This implies that
the number $N$ of massive neutrinos is bigger or equal to three.
If $N>3$, in the flavor basis there are $N_s=N-3$
sterile neutrinos,
$\nu_{s_1}$, \ldots, $\nu_{s_{N_s}}$.
In this case the flavor index $\alpha$ in Eq.~(\ref{mixing})
takes the values
$e,\mu,\tau,s_1,\ldots,s_{N_s}$.

\section{The necessity of at least three independent
$\Delta{\lowercase{\mathbf{m}}}^2$'s}
\label{necessity}

The three evidences in favor of neutrino oscillations
found in solar and atmospheric neutrino experiments
and in the accelerator LSND experiment
imply the existence of at least three independent
neutrino mass-squared differences.
This can be seen by considering
the general expression for the probability of
$\nu_\alpha\to\nu_\beta$
transitions in vacuum,
that can be written as
\cite{Bilenky-Pontecorvo-78,reviews,BGG-98-review}
\begin{equation}
P_{\nu_\alpha\to\nu_\beta}
=
\left|
\sum_{k=1}^{N}
U_{{\alpha}k}^* \,
U_{{\beta}k} \,
\exp\left( - i \, \frac{ \Delta{m}^2_{kj} \, L }{ 2 \, E } \right)
\right|^2
\,,
\label{Posc}
\end{equation}
where
$ \Delta{m}^2_{kj} \equiv m_k^2-m_j^2 $,
$j$ is any of the mass-eigenstate indices,
$L$ is the distance between the neutrino source and detector
and $E$ is the neutrino energy.
The range of $L/E$ characteristic of each type of experiment is different:
$ L / E \sim 10^{11} - 10^{12} \, \mathrm{eV}^{-2} $
for solar neutrino experiments,
$ L / E \sim 10^{2} - 10^{3} \, \mathrm{eV}^{-2} $
for atmospheric neutrino experiments
and
$ L / E \sim 1 \, \mathrm{eV}^{-2} $
for the LSND experiment.
From Eq.~(\ref{Posc}) it is clear that neutrino oscillations
are observable in an experiment only if there is at least one
mass-squared difference
$\Delta{m}^2_{kj}$ such that
\begin{equation}
\frac{ \Delta{m}^2_{kj} \, L }{ 2 \, E }
\gtrsim 0.1
\label{cond1}
\end{equation}
(the precise lower bound depends on the sensitivity of the experiment)
in a significant part of the energy and source-detector distance
intervals of the experiment
(if the condition (\ref{cond1}) is not satisfied,
$
P_{\nu_\alpha\to\nu_\beta}
\simeq
\left|
\sum_k
U_{{\alpha}k}^* \,
U_{{\beta}k}
\right|^2
=
\delta_{\alpha\beta}
$).
Since the range of $L/E$ probed by the LSND experiment is the smaller one,
a large mass-squared difference is needed for LSND oscillations:
\begin{equation}
\Delta{m}^2_{\mathrm{LSND}} \gtrsim 10^{-1} \, \mathrm{eV}^2
\,.
\label{dm2-LSND}
\end{equation}
Specifically,
the maximum likelihood analysis of the LSND data
in terms of two-neutrino oscillations
gives \cite{LSND}
\begin{equation}
0.20 \, \mathrm{eV}^2
\lesssim
\Delta{m}^2_{\mathrm{LSND}}
\lesssim
2.0 \, \mathrm{eV}^2
\,.
\label{LSND-range}
\end{equation}

Furthermore,
from Eq.~(\ref{Posc}) it is clear that 
a dependence of the oscillation probability
from the neutrino energy $E$
and the source-detector distance $L$
is observable only if there is at least one
mass-squared difference
$\Delta{m}^2_{kj}$ such that
\begin{equation}
\frac{ \Delta{m}^2_{kj} \, L }{ 2 \, E }
\sim 1
\,.
\label{cond2}
\end{equation}
Indeed,
all the phases
$ \Delta{m}^2_{kj} L / 2 E \gg 1 $
are washed out by the average over the energy and source-detector
ranges characteristic of the experiment.
Since
a variation of the oscillation probability as a function of neutrino energy
has been observed both in solar and atmospheric neutrino experiments
and the ranges of $L/E$ characteristic
of these two types of experiments are different from each other
and different from the LSND range,
two more mass-squared differences with different scales are needed:
\begin{eqnarray}
&&
\Delta{m}^2_{\mathrm{sun}} \sim 10^{-12} - 10^{-11} \, \mathrm{eV}^2
\qquad
\mbox{(VO)}
\,,
\label{dm2-sun}
\\
&&
\Delta{m}^2_{\mathrm{atm}} \sim 10^{-3} - 10^{-2} \, \mathrm{eV}^2
\,.
\label{dm2-atm}
\end{eqnarray}
The condition (\ref{dm2-sun}) for the solar mass-squared difference
$\Delta{m}^2_{\mathrm{sun}}$
has been obtained under the assumption of vacuum oscillations (VO). 
If the disappearance of solar $\nu_e$'s is due to the MSW effect \cite{MSW},
the condition
\begin{equation}
\Delta{m}^2_{\mathrm{sun}} \lesssim 10^{-4} \, \mathrm{eV}^2
\qquad
\mbox{(MSW)}
\label{dm2-sun-MSW}
\end{equation}
must be fulfilled
in order to have a resonance in the interior of the sun.
Hence,
in the MSW case
$\Delta{m}^2_{\mathrm{sun}}$
must be at least one order of magnitude smaller than
$\Delta{m}^2_{\mathrm{atm}}$.

It is possible to ask if three different scales
of neutrino mass-squared differences are needed even
if the results of the Homestake solar neutrino experiment
is neglected, allowing an energy-independent suppression of the solar $\nu_e$ flux.
The answer is that still the data cannot be fitted with only
two neutrino mass-squared differences
because an energy-independent suppression of the solar $\nu_e$ flux
requires large $\nu_e\to\nu_\mu$ or $\nu_e\to\nu_\tau$
transitions generated by $\Delta{m}^2_{\mathrm{atm}}$
or $\Delta{m}^2_{\mathrm{LSND}}$.
These transitions
are forbidden by the results of the Bugey \cite{Bugey-95}
and CHOOZ \cite{CHOOZ-99}
reactor $\bar\nu_e$ disappearance experiments
and by the non-observation of an up-down asymmetry
of $e$-like events in the Super-Kamiokande
atmospheric neutrino experiment \cite{SK-atm}.

\section{Four-neutrino schemes}
\label{Four-neutrino schemes}

The existence of three different scales of $\Delta{m}^2$
imply that at least four light massive neutrinos must exist in nature.
Here we consider the schemes with four light and mixed neutrinos
\cite{four-models,four-phenomenology,BGKP-96,%
BGG-AB-96,BGG-bounds,BGG-CP,%
Okada-Yasuda-97,BGGS-BBN-98,BGG-BBN-conf,%
Barger-variations-98,BGGS-AB-99,%
BGKM-bb-98,BG-bb-98-99,Giunti-99-bb,BGGKP-bb-99,%
DGKK-99},
which constitute the minimal possibility
that allows to explain all the existing data with neutrino oscillations.
In this case,
in the flavor basis the three active neutrinos $\nu_e$, $\nu_\mu$, $\nu_\tau$
are accompanied by a sterile neutrino $\nu_s$.
Let us notice that the existence of four light massive neutrinos
is favored also
by the possibility that active-sterile neutrino transitions
in neutrino-heated supernova ejecta
could enable the production of $r$-process nuclei
\cite{Fuller-99}.

\begin{figure}[t]
\begin{center}
\includegraphics[bb=13 668 522 827,width=0.99\linewidth]{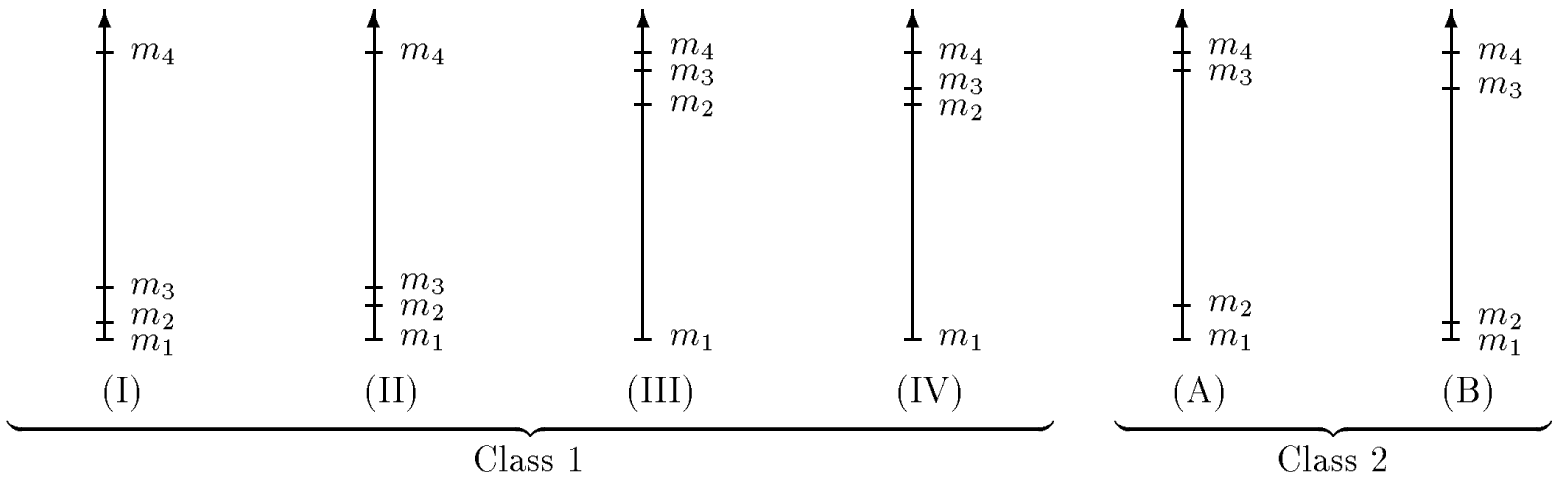}
\refstepcounter{figure}
\label{4spectra}
\small
Figure \ref{4spectra}
\end{center}
\end{figure}

The six types of four-neutrino mass spectra
with
three different scales of $\Delta{m}^2$
that can accommodate
the hierarchy
$
\Delta{m}^2_{\mathrm{sun}}
\ll
\Delta{m}^2_{\mathrm{atm}}
\ll
\Delta{m}^2_{\mathrm{LSND}}
$
are shown qualitatively in Fig.~\ref{4spectra}.
In all these mass spectra there are two groups
of close masses separated by the ``LSND gap'' of the order of 1 eV.
In each scheme the smallest mass-squared
difference corresponds to
$\Delta{m}^2_{\mathrm{sun}}$
($\Delta{m}^2_{21}$ in schemes I and B,
$\Delta{m}^2_{32}$ in schemes II and IV,
$\Delta{m}^2_{43}$ in schemes III and A),
the intermediate one to
$\Delta{m}^2_{\mathrm{atm}}$
($\Delta{m}^2_{31}$ in schemes I and II,
$\Delta{m}^2_{42}$ in schemes III and IV,
$\Delta{m}^2_{21}$ in scheme A,
$\Delta{m}^2_{43}$ in scheme B)
and the largest mass squared difference
$ \Delta{m}^2_{41} = \Delta{m}^2_{\mathrm{LSND}} $
is relevant for the oscillations observed in the LSND experiment.
The six schemes are divided into four schemes of class 1 (I--IV)
in which there is a group of three masses separated from an isolated mass
by the LSND gap,
and two schemes of class 2 (A, B)
in which there are two couples of close masses separated by the LSND gap.

\section{The disfavored schemes of class 1}
\label{class 1}

In the following we will show that
the schemes of class 1 are disfavored by the data
if also the negative
results of short-baseline accelerator and reactor disappearance
neutrino oscillation experiments are taken into account
\cite{BGG-AB-96,Barger-variations-98,BGGS-AB-99}.
Let us remark that in principle one could check which schemes are allowed
by doing a combined fit of all data in the framework of
the most general four-neutrino mixing scheme,
with three mass-squared differences,
six mixing angles and three CP-violating phases
as free parameters.
However,
at the moment it is not possible to perform such a fit
because of the enormous complications due
to the presence of too many parameters
and to the difficulties involved in a combined fit of
the data of different experiments,
which are usually analyzed by the experimental collaborations
using different methods.
Hence,
we think that it is quite remarkable
that one can exclude the schemes of class 1
with the following relatively simple procedure.

Let us define the quantities $d_\alpha$,
with $\alpha=e,\mu,\tau,s$,
in the schemes of class 1 as
\begin{equation}
d_\alpha^{\mathrm{(I)}}
\equiv
|U_{{\alpha}4}|^2
\,,
\qquad
d_\alpha^{\mathrm{(II)}}
\equiv
|U_{{\alpha}4}|^2
\,,
\qquad
d_\alpha^{\mathrm{(III)}}
\equiv
|U_{{\alpha}1}|^2
\,,
\qquad
d_\alpha^{\mathrm{(IV)}}
\equiv
|U_{{\alpha}1}|^2
\,.
\label{dalpha}
\end{equation}
Physically $d_\alpha$ quantifies the mixing of the flavor neutrino $\nu_\alpha$
with the isolated neutrino,
whose mass is separated from the other three by the LSND gap.

The probability of
$\nu_\alpha\to\nu_\beta$
($\beta\neq\alpha$)
and
$\nu_\alpha\to\nu_\alpha$
transitions
(and the corresponding probabilities for antineutrinos)
in short-baseline experiments
are given by \cite{BGG-AB-96,BGG-98-review}
\begin{equation}
P_{\nu_\alpha\to\nu_\beta}
=
A_{\alpha;\beta} \, \sin^2 \frac{ \Delta{m}^2_{41} L }{ 4 E }
\,,
\qquad
P_{\nu_\alpha\to\nu_\alpha}
=
1
-
B_{\alpha;\alpha} \, \sin^2 \frac{ \Delta{m}^2_{41} L }{ 4 E }
\,,
\label{prob}
\end{equation}
with
the oscillation amplitudes
\begin{equation}
A_{\alpha;\beta}
=
4 \, d_\alpha \, d_\beta
\,,
\qquad
B_{\alpha;\alpha}
=
4 \, d_\alpha \, ( 1 - d_\alpha )
\,.
\label{ampli}
\end{equation}
The probabilities (\ref{prob})
have the same form as the corresponding probabilities in the
case of two-neutrino mixing,
$
P_{\nu_\alpha\to\nu_\beta}
=
\sin^2(2\vartheta) \, \sin^2(\Delta{m}^2L/4E)
$
and
$
P_{\nu_\alpha\to\nu_\alpha}
=
1
-
\sin^2(2\vartheta) \, \sin^2(\Delta{m}^2L/4E)
$,
which have been used by all experimental collaborations
for the analysis of the data in order to get information
on the parameters
$\sin^2(2\vartheta)$
and
$\Delta{m}^2$
($\vartheta$ and $\Delta{m}^2$ are, respectively, the mixing angle
and the mass-squared difference in the case of two-neutrino mixing).
Therefore,
we can use the results of their analyses in order to get information
on the corresponding parameters
$A_{\alpha;\beta}$,
$B_{\alpha;\alpha}$
and
$\Delta{m}^2_{41}$.

The exclusion plots obtained in short-baseline
$\bar\nu_e$ and $\nu_\mu$
disappearance experiments
imply that \cite{BGG-AB-96}
\begin{equation}
d_\alpha \leq a^0_\alpha
\qquad \mbox{or} \qquad
d_\alpha \geq 1-a^0_\alpha
\qquad
(\alpha=e,\mu)
\,,
\label{d-small-large}
\end{equation}
with
\begin{equation}
a_\alpha^0
=
\frac{1}{2}
\left( 1 - \sqrt{ 1 - B_{\alpha;\alpha}^0 } \, \right)
\qquad
(\alpha=e,\mu)
\,,
\label{aa0}
\end{equation}
where
$B_{e;e}^0$
and
$B_{\mu;\mu}^0$
are the upper bounds,
that depend on
$\Delta{m}^2_{41}$,
of the oscillation amplitudes
$B_{e;e}$
and
$B_{\mu;\mu}$
given by the exclusion plots of $\bar\nu_e$ and $\nu_\mu$
disappearance experiments.
From the exclusion curves of the Bugey reactor $\bar\nu_e$
disappearance experiment \cite{Bugey-95}
and of the CDHS and CCFR accelerator $\nu_\mu$ disappearance experiments
\cite{CDHS-CCFR-84}
it follows that
$ a_e^0 \lesssim 3 \times 10^{-2} $
for
$\Delta m^2_{41}=\Delta{m}^2_{\mathrm{LSND}}$
in the LSND range (\ref{LSND-range})
and
$ a_\mu^0 \lesssim 0.2 $
for
$\Delta m^2_{41} \gtrsim 0.4 \, \mathrm{eV}^2$
\cite{BGG-98-review}.

\begin{figure}[t!]
\begin{tabular*}{\linewidth}{@{\extracolsep{\fill}}cc}
\begin{minipage}{0.47\linewidth}
\begin{center}
\includegraphics[bb=60 325 511 756,width=0.99\linewidth]{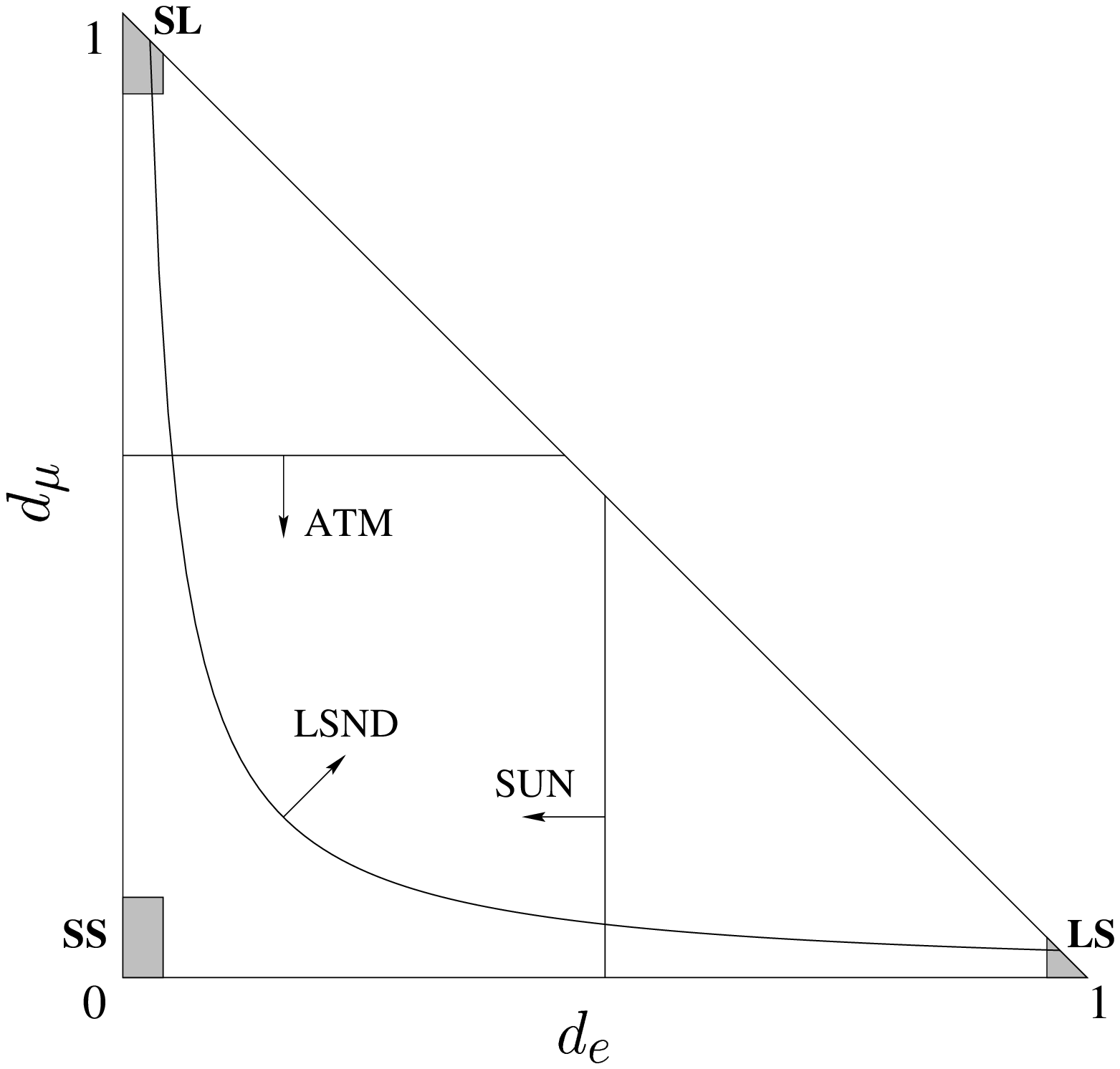}
\refstepcounter{figure}
\label{3dis1}
\small
Figure \ref{3dis1}
\end{center}
\end{minipage}
&
\begin{minipage}{0.47\linewidth}
\begin{center}
\includegraphics[bb=60 326 512 760,width=0.99\linewidth]{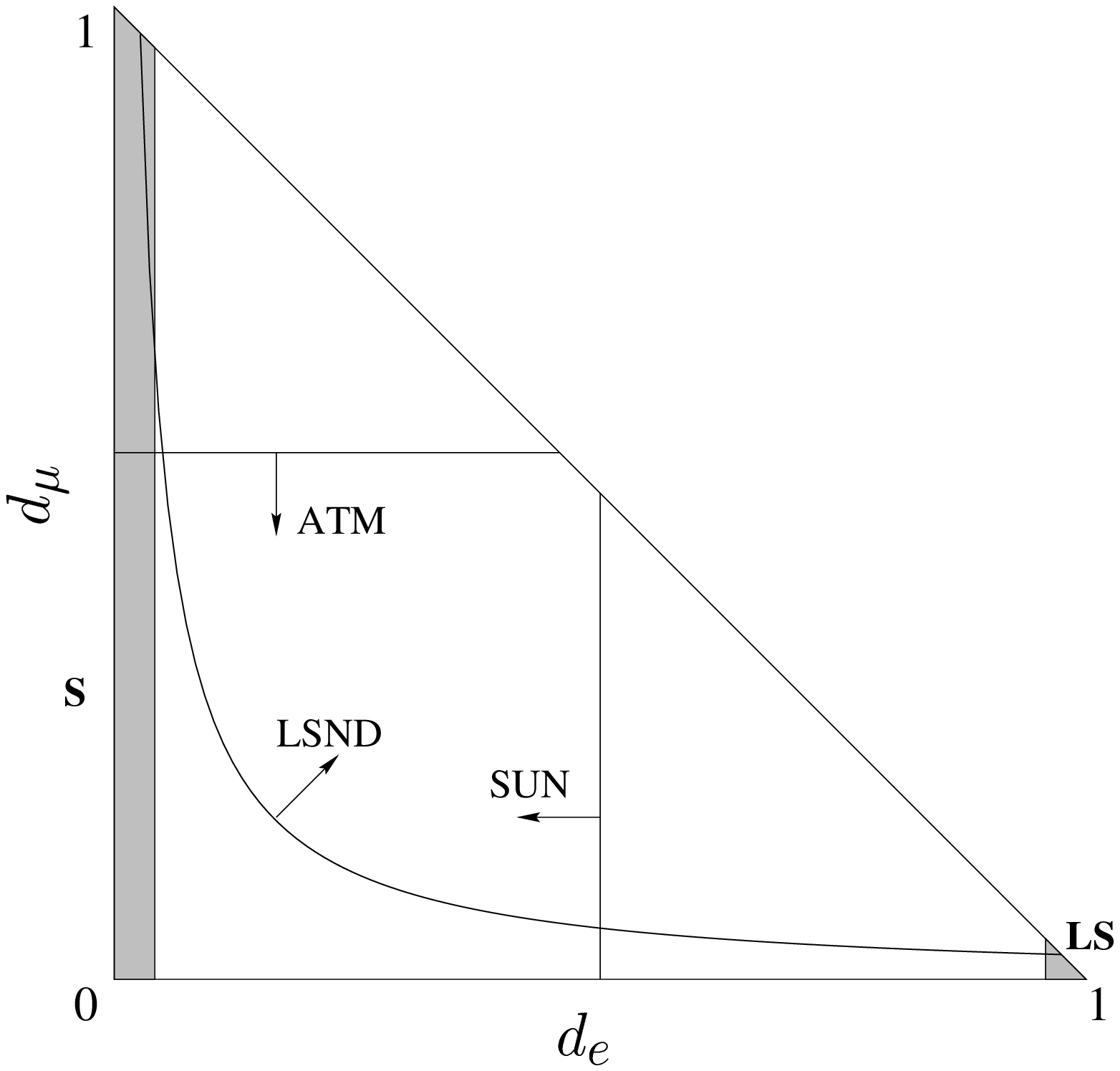}
\refstepcounter{figure}
\label{3dis2}
\small
Figure \ref{3dis2}
\end{center}
\end{minipage}
\end{tabular*}
\end{figure}

Therefore,
the negative results of
short-baseline
$\bar\nu_e$ and $\nu_\mu$
disappearance experiments
imply that $d_e$ and $d_\mu$
are either small or large (close to one).
Taking into account the unitarity limit
$d_e + d_\mu \leq 1$,
for each value of
$\Delta m^2_{41}$
above about
$0.3 \, \mathrm{eV}^2$
there are three regions in the 
$d_e$--$d_\mu$ plane that are allowed by the results of disappearance
experiments:
region SS with small $d_e$ and $d_\mu$,
region LS with large $d_e$ and small $d_\mu$
and
region SL with small $d_e$ and large $d_\mu$.
These three regions are illustrated qualitatively
by the three shadowed areas in Fig.~\ref{3dis1}.
For $\Delta m^2_{41} \lesssim 0.3 \, \mathrm{eV}^2$
there is no constraint on the value of $d_\mu$
from the results of short-baseline
$\nu_\mu$
disappearance experiments and there are two 
regions in the 
$d_e$--$d_\mu$ plane that are allowed by the results of
$\bar\nu_e$ disappearance
experiments:
region S with small $d_e$
and
region LS with large $d_e$ and small $d_\mu$
(the smallness of $d_\mu$ follows from the unitarity bound
$d_e + d_\mu \leq 1$).
These two regions are illustrated qualitatively
by the two shadowed areas in Fig.~\ref{3dis2}.

Let us consider now the results of solar neutrino experiments,
which imply a disappearance of electron neutrinos.
The survival probability of solar $\nu_e$'s
averaged over the fast unobservable oscillations due to
$\Delta{m}^2_{41}$ and $\Delta{m}^2_{31}$
is bounded by
\cite{BGG-AB-96,BGG-98-review}
\begin{equation}
P^{\mathrm{sun}}_{\nu_e\to\nu_e} \geq d_e^2
\,.
\label{solar-bound}
\end{equation}
Therefore, only the possibility
\begin{equation}
d_e \leq a^0_e
\label{de-bound}
\end{equation}
is acceptable in order to explain the observed deficit of solar $\nu_e$'s
with neutrino oscillations.
Indeed, the solar neutrino data imply an upper bound for $d_e$,
that is shown qualitatively by the vertical lines in
Figs. \ref{3dis1} and \ref{3dis2}.
It is clear that the regions LS in Figs. \ref{3dis1} and \ref{3dis2}
are disfavored by the results of solar neutrino experiments.

In a similar way,
since the survival probability of atmospheric $\nu_\mu$'s and $\bar\nu_\mu$'s
is bounded by
\cite{BGG-AB-96,BGG-98-review} 
\begin{equation}
P^{\mathrm{atm}}_{\nu_\mu\to\nu_\mu} \geq d_\mu^2
\,,
\label{atm-bound}
\end{equation}
large values of
$d_\mu$
are incompatible with the asymmetry (\ref{Amu})
observed in the Super-Kamiokande experiment.
The upper bound for $d_\mu$ that follows from atmospheric neutrino data is
shown qualitatively by the horizontal lines in
Figs. \ref{3dis1} and \ref{3dis2}.
It is clear that the region SL in Fig.~\ref{3dis1},
that is allowed
by the results of $\nu_\mu$ short-baseline disappearance experiments
for
$\Delta m^2_{41} \gtrsim 0.3 \, \mathrm{eV}^2$,
and the large--$d_\mu$ part of the region S in Fig.~\ref{3dis2}
are disfavored by the results of atmospheric neutrino experiments.

Therefore,
only the region SS in Fig.~\ref{3dis1}
and the small--$d_\mu$ part of the region S in Fig.~\ref{3dis2}
are allowed by the results of solar and atmospheric neutrino experiments.
In both cases $d_\mu$ is small.
But such small values of $d_\mu$ are disfavored by the results of the LSND
experiment,
that
imply a lower bound
$A^\mathrm{min}_{\mu;e}$
for the amplitude
$A_{\mu;e} = 4 d_e d_\mu$
of $\nu_\mu\to\nu_e$ oscillations.
Indeed, we have
\begin{equation}
d_e \, d_\mu
\geq
A^\mathrm{min}_{\mu;e} / 4
\,.
\label{LSND-lower-bound}
\end{equation}
This bound,
shown qualitatively by the LSND exclusion curves
in Figs. \ref{3dis1} and \ref{3dis2},
excludes region SS in Fig.~\ref{3dis1}
and the small-$d_\mu$ part of region S in Fig.~\ref{3dis2}.
From Figs. \ref{3dis1} and \ref{3dis2}
one can see in a qualitative way that in the schemes of class 1
the results of the solar, atmospheric and LSND experiments
are incompatible with the negative results
of short-baseline experiments.

\begin{figure}[t!]
\begin{tabular*}{\linewidth}{@{\extracolsep{\fill}}cc}
\begin{minipage}{0.47\linewidth}
\begin{center}
\includegraphics[bb=60 148 483 562,width=0.99\linewidth]{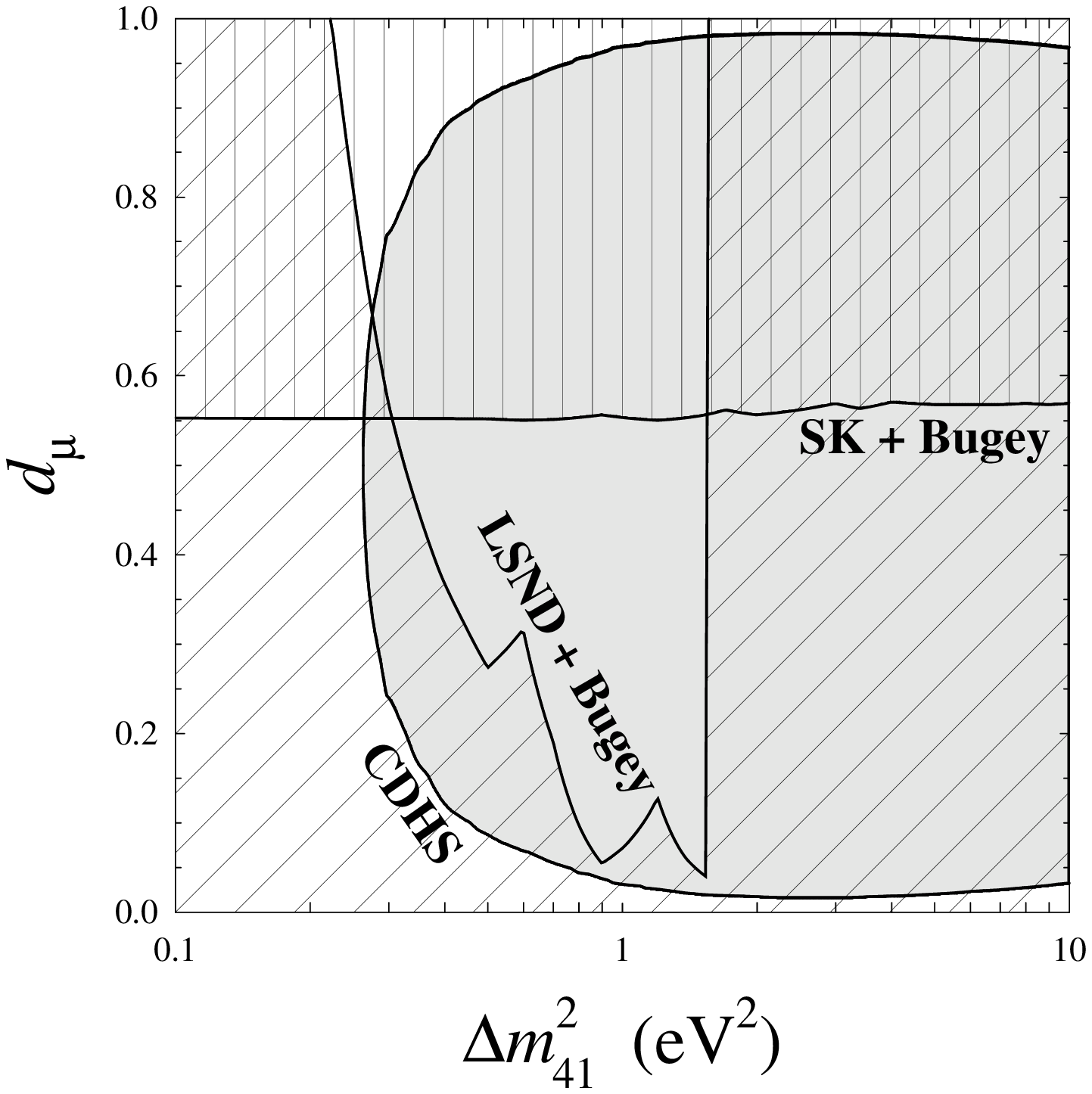}
\refstepcounter{figure}
\label{dmu}
\small
Figure \ref{dmu}
\end{center}
\end{minipage}
&
\begin{minipage}{0.47\linewidth}
\begin{center}
\includegraphics[bb=65 153 485 563,width=0.99\linewidth]{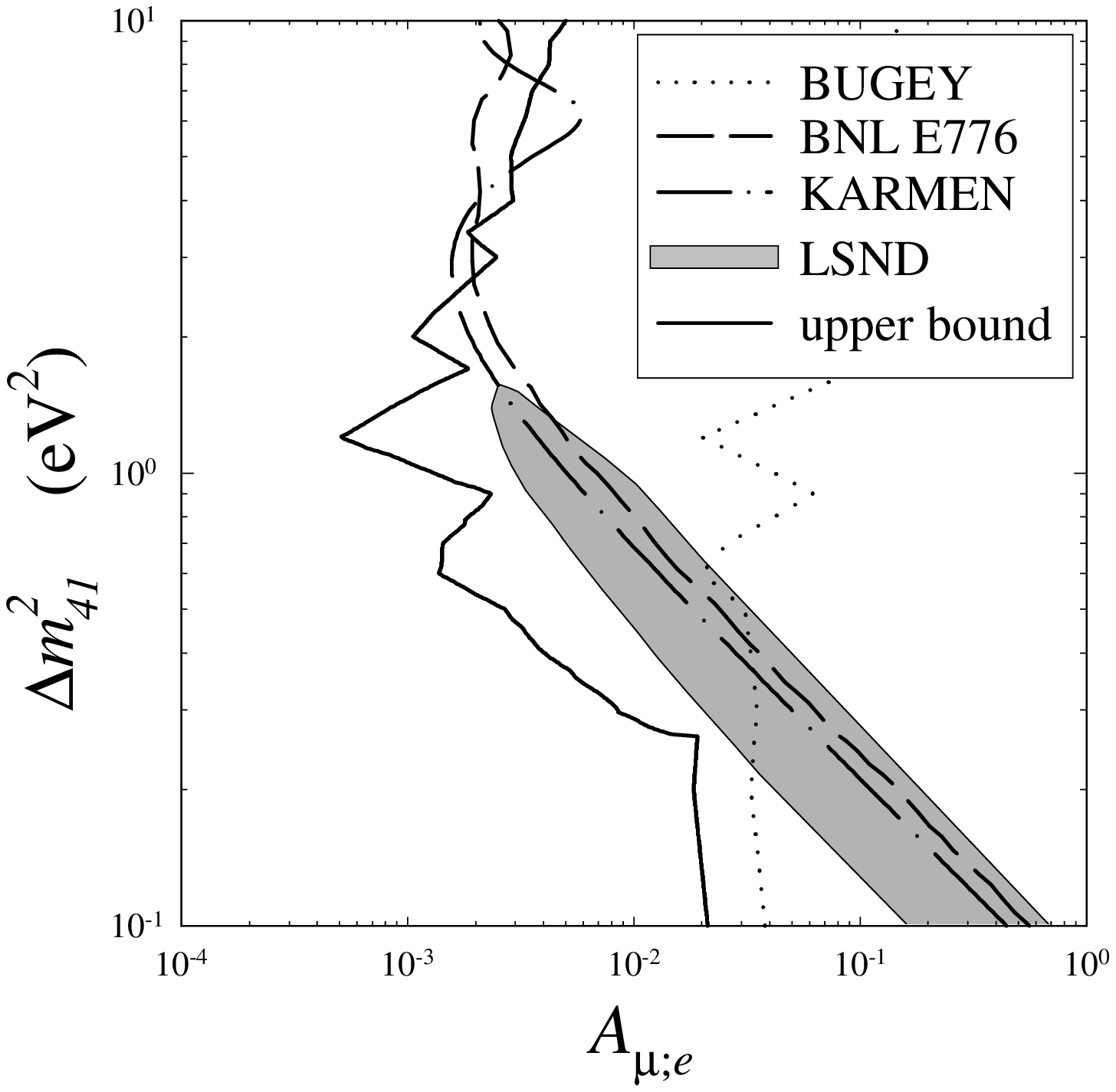}
\refstepcounter{figure}
\label{amuel}
\small
Figure \ref{amuel}
\end{center}
\end{minipage}
\end{tabular*}
\end{figure}

A quantitative illustration of this incompatibility is given
in Fig.~\ref{dmu},
in which the shadowed area is excluded by
the bound
$d_\mu \leq a^0_\mu$
or
$d_\mu \geq 1-a^0_\mu$
obtained from the exclusion plot of the short-baseline
CDHS
$\nu_\mu$ disappearance experiment.
The horizontal line in Fig.~\ref{dmu} represents the upper bound
\begin{equation}
d_\mu \lesssim 0.55 \equiv a^{\mathrm{SK}}_\mu
\label{dmu-bound}
\end{equation}
(the vertically hatched area above the line is excluded)
obtained from
the Super-Kamiokande asymmetry (\ref{Amu})
and
the exclusion curve of the Bugey $\bar\nu_e$
disappearance experiment
\cite{BGGS-AB-99}.
It is clear that the results of short-baseline disappearance experiments
and the Super-Kamiokande asymmetry (\ref{Amu})
imply that
$ d_\mu \lesssim 0.55 $
for
$ \Delta{m}^2_{41} \lesssim 0.3 \, \mathrm{eV}^2 $
and that
$ d_\mu $
is very small for
$ \Delta{m}^2_{41} \gtrsim 0.3 \, \mathrm{eV}^2 $.
These small values of $ d_\mu $ are disfavored by
the curve in Fig.~\ref{dmu} labelled
LSND + Bugey,
which represents the constraint
\begin{equation}\label{LSND}
d_\mu \geq A^\mathrm{min}_{\mu;e}/4a^0_e
\end{equation}
(the diagonally hatched area is excluded),
derived from the inequality (\ref{LSND-lower-bound})
using the bound (\ref{de-bound}).
Hence,
in the framework of the schemes of class 1
there is no range of $d_\mu$
that is compatible with all the experimental data.

Another way for seeing the
incompatibility of the experimental results with the schemes of class 1
is presented in Fig.~\ref{amuel},
where we have plotted in the $A_{\mu;e}$--$\Delta{m}^2_{41}$ plane
the upper bound
$ A_{\mu;e} \leq 4 \, a^0_e \, a^0_\mu $
for
$ \Delta{m}^2_{41} > 0.26 \, \mathrm{eV}^2 $
and
$ A_{\mu;e} \leq 4 \, a^0_e \, a^{\mathrm{SK}}_\mu $
for
$ \Delta{m}^2_{41} < 0.26 \, \mathrm{eV}^2 $
(solid line, the region on the right is excluded).
One can see that this constraint is incompatible with
the LSND-allowed region
(shadowed area).

Summarizing,
we have reached the conclusion that the four schemes of class 1
shown in Fig.~\ref{4spectra} are disfavored by the data.

\section{The favored schemes of class 2}
\label{class 2}

\begin{figure}[t!]
\begin{tabular*}{\linewidth}{@{\extracolsep{\fill}}cc}
\begin{minipage}{0.47\linewidth}
\begin{center}
\includegraphics[bb=60 326 510 741,width=0.99\linewidth]{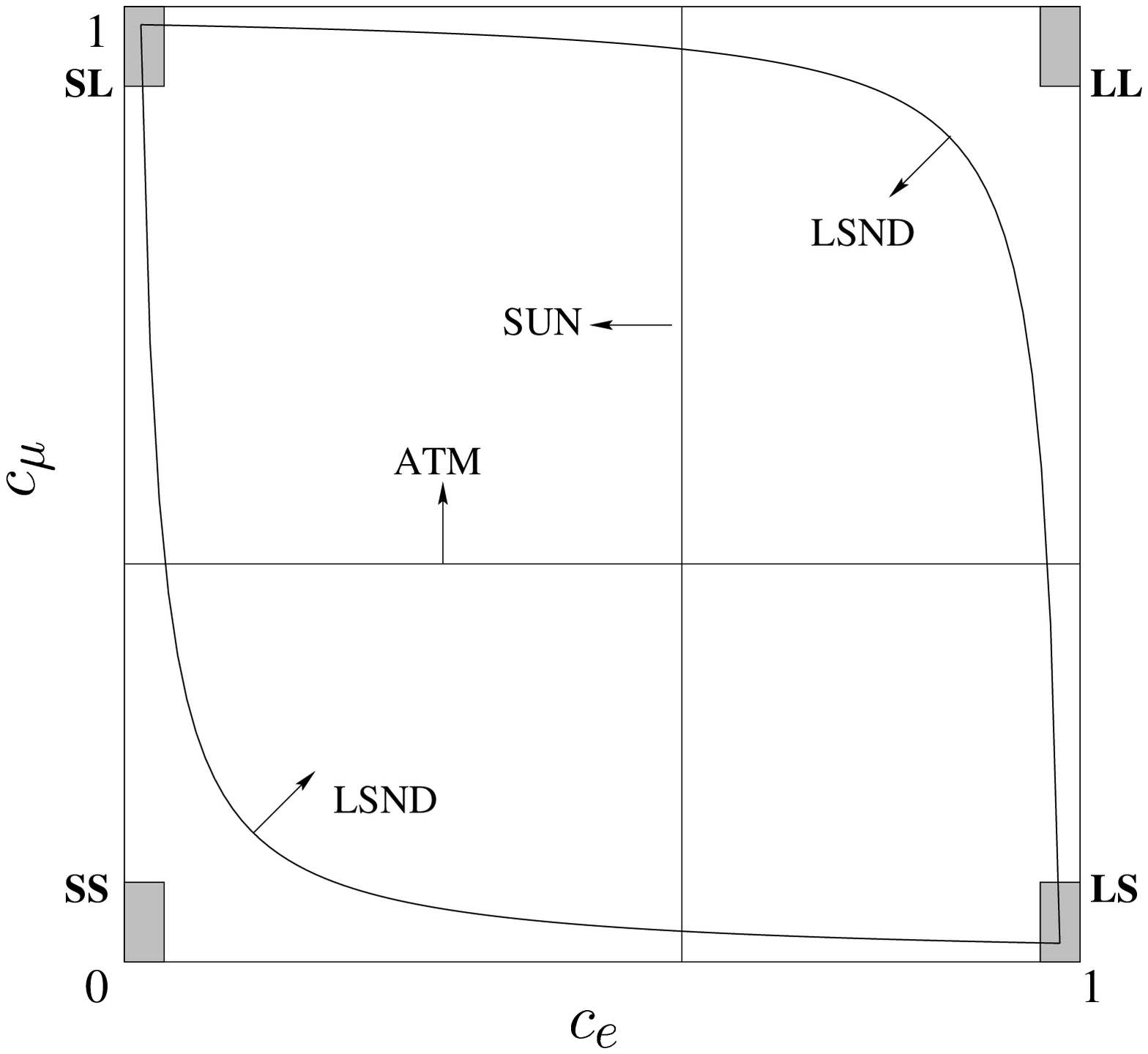}
\refstepcounter{figure}
\label{4dis1}
\small
Figure \ref{4dis1}
\end{center}
\end{minipage}
&
\begin{minipage}{0.47\linewidth}
\begin{center}
\includegraphics[bb=60 326 512 751,width=0.99\linewidth]{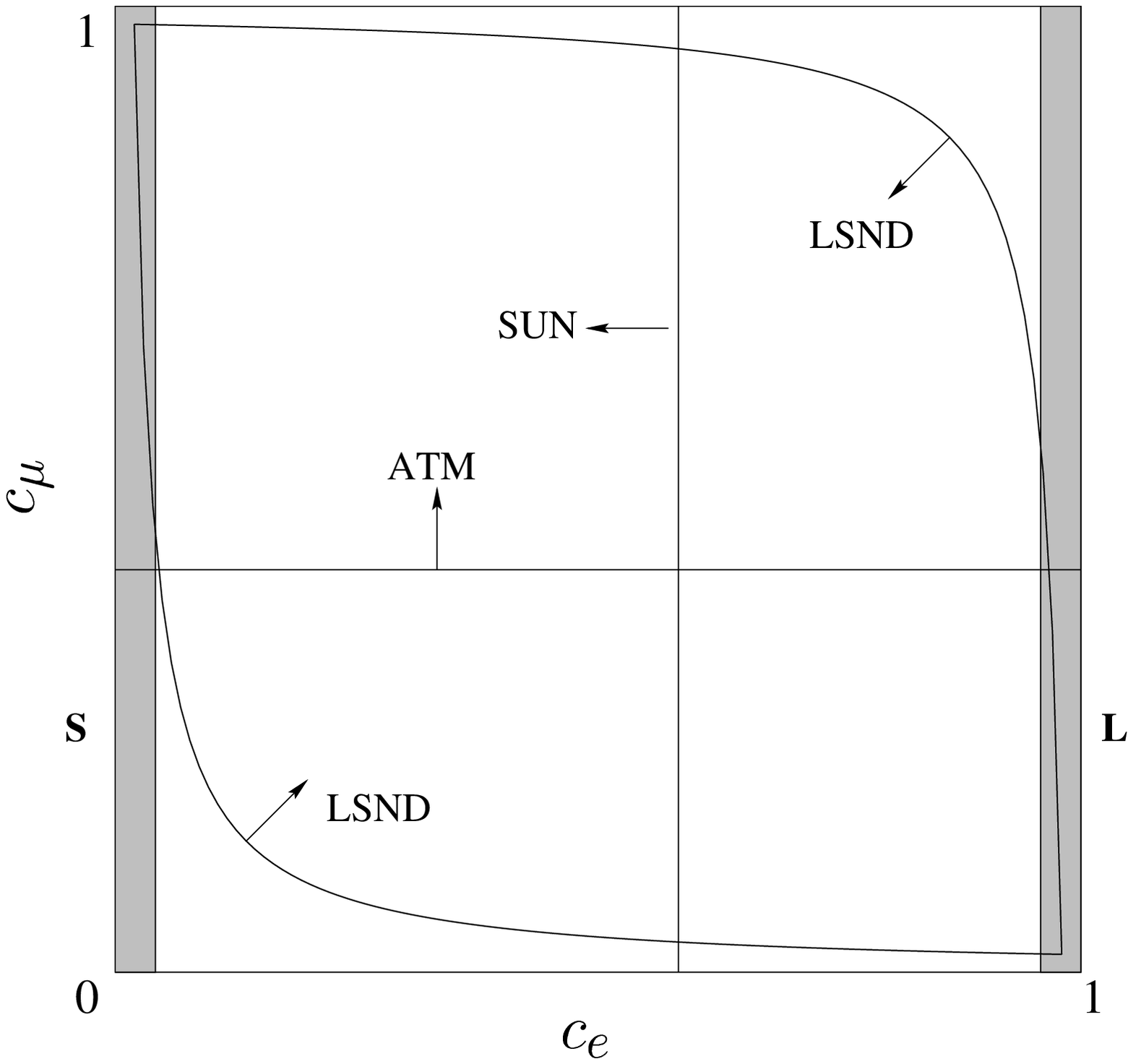}
\refstepcounter{figure}
\label{4dis2}
\small
Figure \ref{4dis2}
\end{center}
\end{minipage}
\end{tabular*}
\end{figure}

The four-neutrino schemes of class 2
are compatible with the results of all neutrino oscillation experiments
if
the mixing of $\nu_e$ with the two mass eigenstates responsible
for the oscillations of solar neutrinos
($\nu_3$ and $\nu_4$ in scheme A
and
$\nu_1$ and $\nu_2$ in scheme B)
is large
and
the mixing of $\nu_\mu$ with the two mass eigenstates responsible
for the oscillations of atmospheric neutrinos
($\nu_1$ and $\nu_2$ in scheme A
and
$\nu_3$ and $\nu_4$ in scheme B)
is large
\cite{BGKP-96,BGG-AB-96,Barger-variations-98,BGGS-AB-99}.
This is illustrated qualitatively in Figs. \ref{4dis1} and \ref{4dis2},
as we are going to explain.

Let us define the quantities $c_\alpha$,
with $\alpha=e,\mu,\tau,s$,
in the schemes A and B as
\begin{equation}
c_\alpha^{\mathrm{(A)}}
\equiv
\sum_{k=1,2}
|U_{{\alpha}k}|^2
\,,
\qquad
c_\alpha^{\mathrm{(B)}}
\equiv
\sum_{k=3,4}
|U_{{\alpha}k}|^2
\,.
\label{04}
\end{equation}
Physically $c_\alpha$ quantify the mixing of the flavor neutrino $\nu_\alpha$
with the two massive neutrinos whose $\Delta{m}^2$
is relevant for the oscillations of atmospheric neutrinos
($\nu_1$, $\nu_2$ in scheme A
and
$\nu_3$, $\nu_4$ in scheme B).
The exclusion plots obtained in short-baseline
$\bar\nu_e$ and $\nu_\mu$
disappearance experiments
imply that \cite{BGG-AB-96}
\begin{equation}
c_\alpha \leq a^0_\alpha
\qquad \mbox{or} \qquad
c_\alpha \geq 1-a^0_\alpha
\qquad
(\alpha=e,\mu)
\,,
\label{c-small-large}
\end{equation}
with $a_\alpha^0$ given in Eq.~(\ref{aa0}).

The shadowed areas in Figs. \ref{4dis1} and \ref{4dis2}
illustrate qualitatively
the regions in the
$c_e$--$c_\mu$ plane
allowed by the negative results of short-baseline
$\bar\nu_e$ and $\nu_\mu$
disappearance experiments.
Figure~\ref{4dis1} is valid for
$\Delta m^2_{41} \gtrsim 0.3 \, \mathrm{eV}^2$
and shows that there are four regions
allowed by the results of short-baseline disappearance experiments:
region SS with small $c_e$ and $c_\mu$,
region LS with large $c_e$ and small $c_\mu$,
region SL with small $c_e$ and large $c_\mu$
and
region LL with large $c_e$ and $c_\mu$.
The quantities $c_e$ and $c_\mu$
can be both large,
because the unitarity of the mixing matrix
imply that
$c_\alpha + c_\beta \leq 2$
and
$0 \leq c_\alpha \leq 1$
for $\alpha,\beta=e,\mu,\tau,s$.
Figure~\ref{4dis2} is valid for
$\Delta m^2_{41} \lesssim 0.3 \, \mathrm{eV}^2$,
where there is no constraint on the value of $c_\mu$
from the results of short-baseline
$\nu_\mu$
disappearance experiments.
It shows that there are two regions
allowed by the results of short-baseline $\bar\nu_e$
disappearance experiments:
region S with small $c_e$
and
region L with large $c_e$.

Let us take now into account the results of solar neutrino experiments.
Large values of $c_e$ are incompatible with solar neutrino oscillations
because in this case $\nu_e$ has large mixing with
the two massive neutrinos responsible for atmospheric neutrino oscillations
and, through the unitarity of the mixing matrix,
small mixing with
the two massive neutrinos responsible for solar neutrino oscillations.
Indeed,
in the schemes of class 2
the survival probability
$P^{\mathrm{sun}}_{\nu_e\to\nu_e}$
of solar $\nu_e$'s is bounded by
\cite{BGG-AB-96,BGG-98-review}
\begin{equation}
P^{\mathrm{sun}}_{\nu_e\to\nu_e}
\geq
c_e^2 / 2
\,,
\label{c-solar-bound-1}
\end{equation}
and its possible variation
$\Delta P^{\mathrm{sun}}_{\nu_e\to\nu_e}(E)$
with neutrino energy $E$
is limited by 
\cite{BGG-AB-96,BGG-98-review}
\begin{equation}
\Delta P^{\mathrm{sun}}_{\nu_e\to\nu_e}(E)
\leq
\left( 1 - c_e \right)^2
\,.
\label{c-solar-bound-2}
\end{equation}
If $c_e$ is large as in the LS or LL regions of Fig.~\ref{4dis1}
or in the L region of Fig.~\ref{4dis2},
we have
\begin{equation}
P^{\mathrm{sun}}_{\nu_e\to\nu_e}
\geq
\frac{ \left( 1 - a^0_e \right)^2 }{ 2 }
\simeq
\frac{ 1 }{ 2 }
\,,
\qquad
\Delta P^{\mathrm{sun}}_{\nu_e\to\nu_e}(E)
\leq
(a^0_e)^2
\lesssim
10^{-3}
\,,
\label{c-solar-bound-3}
\end{equation}
for
$\Delta m^2_{41}=\Delta{m}^2_{\mathrm{LSND}}$
in the LSND range (\ref{LSND-range}).
Therefore
$P^{\mathrm{sun}}_{\nu_e\to\nu_e}$
is bigger than 1/2 and
practically
does not depend on neutrino energy.
Since this is incompatible with the results of solar neutrino experiments
interpreted in terms of neutrino oscillations
\cite{BGG-98-review,sun-analysis},
we conclude that the regions
LS and LL in Fig.~\ref{4dis1}
and the region L in Fig.~\ref{4dis2}
are disfavored by solar neutrino data,
as illustrated qualitatively by the vertical exclusion lines in
Figs. \ref{4dis1} and \ref{4dis2}.

Let us consider now the results of atmospheric neutrino experiments.
Small values of $c_\mu$ are incompatible with atmospheric neutrino oscillations
because in this case $\nu_\mu$ has
small mixing with
the two massive neutrinos responsible for atmospheric neutrino oscillations.
Indeed, the survival probability of
atmospheric $\nu_\mu$'s
is bounded by
\cite{BGG-AB-96,BGG-98-review}
\begin{equation}
P^{\mathrm{atm}}_{\nu_\mu\to\nu_\mu} \geq \left( 1 - c_\mu \right)^2
\,,
\label{c-atm-bound-1}
\end{equation}
and it can be shown
\cite{BGGS-AB-99}
that the Super-Kamiokande asymmetry (\ref{Amu})
and
the exclusion curve of the Bugey $\bar\nu_e$
disappearance experiment
imply the upper bound
\begin{equation}
c_\mu \gtrsim 0.45 \equiv b^{\mathrm{SK}}_\mu
\,.
\label{c-atm-bound-2}
\end{equation}
This limit is depicted qualitatively by the horizontal
exclusion lines in Figs. \ref{4dis1} and \ref{4dis2}.
Therefore,
we conclude that the regions
SS and LS in Fig.~\ref{4dis1}
and the small-$c_\mu$ parts
of the regions S and L in Fig.~\ref{4dis2}
are disfavored by atmospheric neutrino data.

Finally, let us consider the results of the LSND experiment.
In the schemes of class 2 the amplitude of
short-baseline $\nu_\mu\to\nu_e$ oscillations is given by
\begin{equation}
A_{\mu;e}
=
\bigg|
\sum_{k=1,2} U_{ek} U_{\mu k}^*
\bigg|^2
=
\bigg|
\sum_{k=3,4} U_{ek} U_{\mu k}^*
\bigg|^2
\,.
\label{Amuel}
\end{equation}
The second equality in Eq. (\ref{Amuel}) is due to the unitarity of the mixing
matrix.
Using the Cauchy--Schwarz inequality we obtain
\begin{equation}
c_e \, c_\mu
\geq
A^\mathrm{min}_{\mu;e} / 4
\qquad
\mbox{and}
\qquad
\left( 1 - c_e \right) \left( 1 - c_\mu \right)
\geq
A^\mathrm{min}_{\mu;e} / 4
\,,
\label{Amuel-bounds}
\end{equation}
where
$A^\mathrm{min}_{\mu;e}$
is the minimum value of the oscillation amplitude
$A_{\mu;e}$
observed in the LSND experiment.
The bounds (\ref{Amuel-bounds})
are illustrated qualitatively in Figs. \ref{4dis1} and \ref{4dis2}.
One can see that
the results of the LSND experiment confirm the exclusion of
the regions
SS and LL in Fig.~\ref{4dis1}
and the exclusion
of the small-$c_\mu$ part of region S
and
of the large-$c_\mu$ part of region L
in Fig.~\ref{4dis2}.

Summarizing,
if
$ \Delta{m}^2_{41} \gtrsim 0.3 \, \mathrm{eV}^2 $
only the region SL in Fig.~\ref{4dis1},
with
\begin{equation}
c_e \leq a^0_e
\qquad \mbox{and} \qquad
c_\mu \geq 1 - a^0_\mu
\,,
\label{c-bounds-1}
\end{equation}
is compatible with the results of all neutrino oscillation experiments.
If
$ \Delta{m}^2_{41} \lesssim 0.3 \, \mathrm{eV}^2 $
only the large-$c_\mu$ part of region S in Fig.~\ref{4dis2},
with
\begin{equation}
c_e \leq a^0_e
\qquad \mbox{and} \qquad
c_\mu \geq b^{\mathrm{SK}}_\mu
\,,
\label{c-bounds-2}
\end{equation}
is compatible with the results of all neutrino oscillation experiments.
Therefore,
in any case $c_e$ is small and $c_\mu$ is large.
However,
it is important to notice that,
as shown clearly in Figs. \ref{4dis1} and \ref{4dis2},
the inequalities
(\ref{Amuel-bounds})
following from the LSND observation of
short-baseline $\nu_\mu\to\nu_e$ oscillations
imply that $c_e$, albeit small,
has a lower bound and
$c_\mu$, albeit large,
has an upper bound:
\begin{equation}
c_e \gtrsim A^\mathrm{min}_{\mu;e} / 4
\qquad \mbox{and} \qquad
c_\mu \lesssim 1 - A^\mathrm{min}_{\mu;e} / 4
\,.
\label{c-bounds-3}
\end{equation}

\section{Conclusions}
\label{Conclusions}

We have seen that only the two four-neutrino schemes A and B
of class 2 in Fig.~\ref{4spectra}
are compatible with the results of all neutrino oscillation experiments.
Furthermore,
we have shown that the quantities $c_e$ and $c_\mu$
in these two schemes must be, respectively,
small and large.
Physically $c_\alpha$,
defined in Eq.~(\ref{04}),
quantify the mixing of the flavor neutrino $\nu_\alpha$
with the two massive neutrinos whose $\Delta{m}^2$
is relevant for the oscillations of atmospheric neutrinos
($\nu_1$, $\nu_2$ in scheme A
and
$\nu_3$, $\nu_4$ in scheme B).

The smallness of $c_e$ implies that
electron neutrinos do not oscillate
in atmospheric and long-baseline neutrino oscillation experiments.
Indeed,
one can obtain rather stringent upper bounds
for the probability of $\nu_e$ transitions into any other state
\cite{BGG-bounds}
and for the size of CP or T violation that could be measured
in long-baseline experiments
in the $\nu_\mu\leftrightarrows\nu_e$
and $\bar\nu_\mu\leftrightarrows\bar\nu_e$ channels
\cite{BGG-CP}.

Let us consider
\cite{BGKM-bb-98,BG-bb-98-99,Giunti-99-bb,BGGKP-bb-99}
now
the effective Majorana mass in neutrinoless double-$\beta$ decay,
\begin{equation}
|\langle{m}\rangle|
=
\bigg| \sum_{k=1}^4 U_{ek}^2 \, m_k \bigg|
\,.
\label{Majorana}
\end{equation}

In scheme A,
since $c_e$ is small,
the effective Majorana mass is approximately given by
\begin{equation}
|\langle{m}\rangle|
\simeq
\left| U_{e3}^2 + U_{e4}^2 \right| m_4
\simeq
\left| U_{e3}^2 + U_{e4}^2 \right| \sqrt{\Delta{m}^2_{\mathrm{LSND}}}
\,.
\label{bb-A}
\end{equation}
Therefore,
in scheme A the effective Majorana mass can be as large as
$\sqrt{\Delta{m}^2_{\mathrm{LSND}}}$.
Since $c_e$ is small,
in scheme A we have
$|U_{e3}|^2 \simeq \cos^2\vartheta_{\mathrm{sun}}$
and
$|U_{e4}|^2 \simeq \sin^2\vartheta_{\mathrm{sun}}$,
where
$\vartheta_{\mathrm{sun}}$
is the mixing angle determined from the two-generation analysis of
solar neutrino data \cite{sun-analysis}.
In the case of the small mixing angle MSW solution of the solar neutrino problem
$|U_{e4}| \ll |U_{e3}| \simeq 1$
and
from Eq.~(\ref{bb-A}) one can see that
$
|\langle{m}\rangle|
\simeq
\sqrt{\Delta{m}^2_{\mathrm{LSND}}}
$
In the case of the large mixing angle MSW solution,
$\sin^2 2\vartheta_{\mathrm{sun}}$
is constrained to be less than about 0.97 at 99\% CL
\cite{concha}
and the effective Majorana mass lies in
the range
\cite{BGGKP-bb-99}
$7 \times 10^{-2} \, \mathrm{eV}
\lesssim |\langle{m}\rangle| \lesssim
1.4 \, \mathrm{eV}$.
In the case of the vacuum oscillation solution,
$\sin^2 2\vartheta_{\mathrm{sun}}$
can be as large as one and
there is no lower bound for $|\langle{m}\rangle|$.
If future measurements will
show the correctness of a large mixing angle solution of the solar 
neutrino problem (due to vacuum oscillations or the MSW effect),
the measurement of
$|\langle{m}\rangle|$
would give information
\cite{BGKP-96,BGGKP-bb-99}
on the value of a
Majorana phase in the mixing matrix $U$,
that does not contribute to neutrino oscillations
\cite{BHP80-Kobzarev80-Doi81-Langacker87}.

In scheme B the contribution of the ``heavy'' neutrino masses
$m_3$ and $m_4$
to the effective Majorana mass is strongly suppressed
\cite{BGKM-bb-98,BG-bb-98-99,Giunti-99-bb,BGGKP-bb-99}:
\begin{equation}
|\langle{m}\rangle|_{34}
\equiv
\left| U_{e3}^2 \, m_3 + U_{e4}^2 \, m_4 \right|
\lesssim
c_e \, m_4
\leq
a^0_e \, \sqrt{\Delta{m}^2_{\mathrm{LSND}}}
\simeq
2 \times 10^{-2} \, \mathrm{eV}
\,.
\label{bb-B}
\end{equation}
Therefore,
if future neutrinoless double-$\beta$ decay experiments
will find that
$|\langle{m}\rangle| \gtrsim 2 \times 10^{-2} \, \mathrm{eV}$,
it would mean that scheme B is excluded, or that
neutrinoless double-$\beta$ decay proceeds through
other mechanisms,
not involving the effective Majorana mass $|\langle{m}\rangle|$. 

Finally,
if the upper bound $ N_\nu^{\mathrm{BBN}} < 4 $
for the effective number of neutrinos
in Big-Bang Nucleosynthesis is correct
\cite{BBN-Nnu},
the mixing of $\nu_s$
with the two mass eigenstates responsible
for the oscillations of atmospheric neutrinos
must be very small \cite{Okada-Yasuda-97,BGGS-BBN-98,BGG-BBN-conf}.
In this case atmospheric neutrinos oscillate only in the
$\nu_\mu\to\nu_\tau$ channel
and solar neutrinos oscillate only in the
$\nu_e\to\nu_s$ channel.
This is very important because it implies that the two-generation
analyses of solar and atmospheric neutrino data
give correct information on neutrino mixing
in the two four-neutrino schemes A and B.
Otherwise, it will be necessary to reanalyze the solar and atmospheric
neutrino data using a general formalism
that takes into account the possibility of simultaneous transitions into
active and sterile neutrinos
in solar and atmospheric neutrino experiments \cite{DGKK-99}.

\section*{Acknowledgments}

I would like to thank S.M. Bilenky for his friendship,
for teaching me a lot of physics,
for many interesting and stimulating discussions
and
for a fruitful collaboration lasting several years.
I would like to thank also W. Grimus and T. Schwetz
for enjoyable collaboration on the topics presented in this report.


\begin{thebibliography}{10}

\bibitem{Pontecorvo-57-58}
B. Pontecorvo,
J. Exptl. Theoret. Phys. \textbf{33}, 549 (1957)
[Sov. Phys. JETP \textbf{6}, 429 (1958)];
B. Pontecorvo,
J. Exptl. Theoret. Phys. \textbf{34}, 247 (1958)
[Sov. Phys. JETP \textbf{7}, 172 (1958)].

\bibitem{Pontecorvo-67}
B. Pontecorvo,
Zh. Eksp. Teor. Fiz. \textbf{53}, 1717 (1967)
[Sov. Phys. JETP \textbf{26}, 984 (1968)].

\bibitem{Davis-68}
R. Davis \textit{et al.}, 
Phys. Rev. Lett. \textbf{20}, 1205 (1968).

\bibitem{Bahcall-68}
J.N. Bahcall \textit{et al.}, 
Phys. Rev. Lett. \textbf{20}, 1209 (1968).

\bibitem{Gribov-Pontecorvo-69}
V. Gribov and B. Pontecorvo,
Phys. Lett. \textbf{B28}, 493 (1969).

\bibitem{Bilenky-Pontecorvo-76a}
S.M. Bilenky and B. Pontecorvo,
Lett. Nuovo Cim. \textbf{17}, 569 (1976).

\bibitem{Bilenky-Pontecorvo-78}
S.M. Bilenky and B. Pontecorvo,
Phys. Rep. \textbf{41}, 225 (1978).

\bibitem{reviews}
S.M. Bilenky and S.T. Petcov,
Rev. Mod. Phys. \textbf{59}, 671 (1987);
C.W. Kim and A. Pevsner,
\textit{Neutrinos in Physics and Astrophysics},
Harwood Academic Press, Chur, Switzerland, 1993;
K. Zuber,
Phys. Rep. \textbf{305}, 295 (1998), hep-ph/9811267;
G. Raffelt,
hep-ph/9902271;
E. Torrente-Lujan,
hep-ph/9902339;
A.B. Balantekin and W.C. Haxton,
nucl-th/9903038;
W.C. Haxton and B.R. Holstein,
hep-ph/9905257;
P. Fisher, B. Kayser and K.S. McFarland,
hep-ph/9906244;
R.D. Peccei,
hep-ph/9906509.

\bibitem{BGG-98-review}
S.M. Bilenky, C. Giunti and W. Grimus,
hep-ph/9812360
[Prog. Part. Nucl. Phys. \textbf{43}, in press].

\bibitem{see-saw}
M. Gell-Mann, P. Ramond and R. Slansky,
in \textit{Supergravity}, p.~315,
1979;
T. Yanagida,
Proc. of the
\textit{Workshop on Unified Theory and the Baryon Number of the Universe},
KEK, Japan, 1979;
R.N. Mohapatra and G. Senjanovi{\'c},
Phys. Rev. Lett. \textbf{44}, 912 (1980).

\bibitem{effective}
S. Weinberg,
Phys. Rev. Lett. \textbf{43}, 1566 (1979);
Phys. Rev. \textbf{D22}, 1694 (1980);
Int. J. Mod. Phys. \textbf{A2}, 301 (1987);
H.A. Weldon and A. Zee,
Nucl. Phys. \textbf{B173}, 269 (1980);
E.Kh. Akhmedov \textit{et al.},
Phys. Rev. Lett. \textbf{69}, 3013 (1992);
Phys. Rev. \textbf{D47}, 3245 (1993).

\bibitem{SK-atm}
Y. Fukuda \textit{et al.} (Super-Kamiokande Coll.),
Phys. Rev. Lett. \textbf{81}, 1562 (1998);
K. Scholberg (Super-Kamiokande Coll.),
hep-ex/9905016.

\bibitem{Lipari-99}
P. Lipari,
hep-ph/9904443.

\bibitem{atm-exp-contained}
Y. Fukuda \textit{et al.} (Kamiokande Coll.),
Phys. Lett. \textbf{B335}, 237 (1994);
R. Becker-Szendy \textit{et al.} (IMB Coll.),
Nucl. Phys. B (Proc. Suppl.) \textbf{38}, 331 (1995);
W.W.M. Allison \textit{et al.} (Soudan Coll.),
Phys. Lett. \textbf{B449}, 137 (1999).

\bibitem{atm-exp-upmu}
Y. Fukuda \textit{et al.} (Super-Kamiokande Coll.),
Phys. Rev. Lett. \textbf{82}, 2644 (1999);
A. Habig (Super-Kamiokande Coll.), hep-ex/9903047;
M. Ambrosio \textit{et al.} (MACRO Coll.),
Phys. Lett. \textbf{B434}, 451 (1998);
P. Bernardini (MACRO Coll.),
hep-ex/9906019.

\bibitem{CHOOZ-99}
M. Apollonio \textit{et al.} (CHOOZ Coll.),
Phys. Lett. \textbf{B420}, 397 (1998);
hep-ex/9907037.

\bibitem{sun-exp}
B.T. Cleveland \textit{et al.},
Astrophys. J. \textbf{496}, 505 (1998);
K.S. Hirata \textit{et al.} (Kamiokande Coll.),
Phys. Rev. Lett. \textbf{77}, 1683 (1996);
W. Hampel \textit{et al.} (GALLEX Coll.),
Phys. Lett. \textbf{B447}, 127 (1999);
J.N. Abdurashitov \textit{et al.} (SAGE Coll.),
astro-ph/9907113;
Y. Fukuda \textit{et al.} (Super-Kamiokande Coll.),
Phys. Rev. Lett. \textbf{81}, 1158 (1998);
Phys. Rev. Lett. \textbf{82}, 2430 (1999);
M.B. Smy (Super-Kamiokande Coll.),
hep-ex/9903034.

\bibitem{LSND}
C. Athanassopoulos \textit{et al.} (LSND Coll.),
Phys. Rev. Lett. \textbf{75}, 2650 (1995);
Phys. Rev. Lett. \textbf{77}, 3082 (1996);
Phys. Rev. Lett. \textbf{81}, 1774 (1998);
G. Mills (LSND Coll.),
Talk presented at the
XXXIV$^{\mathrm{th}}$
Rencontres de Moriond
\textit{Electroweak Interactions and Unified Theories},
Les Arcs, March 1999 
(http://{\-}moriond.{\-}in2p3.{\-}fr/{\-}EW/{\-}transparencies).

\bibitem{BP98}
J.N. Bahcall \textit{et al.}, 
Phys. Lett. \textbf{B433}, 1 (1998). 

\bibitem{sun-analysis}
J.N. Bahcall, P.I. Krastev and A.Yu. Smirnov,
Phys. Rev. \textbf{D58}, 096016 (1998);
Y. Fukuda \textit{et al.},
Phys. Rev. Lett. \textbf{82}, 1810 (1999);
V. Barger and K. Whisnant,
Phys. Lett. \textbf{B456}, 54 (1999);
M.C. Gonzalez-Garcia \textit{et al.},
hep-ph/9906469.

\bibitem{LSND-check}
Booster Neutrino Experiment (BooNE), http://{\-}nu1.lampf.lanl.gov/{\-}BooNE;
I-216 $\nu_\mu\to\nu_e$ proposal at CERN,
http://{\-}chorus01.{\-}cern.ch/\~{}pzucchel/{\-}loi/;
Oak Ridge Large Neutrino Detector, http://{\-}www.{\-}phys.{\-}subr.{\-}edu/{\-}orland/;
NESS: Neutrinos at the European Spallation Source,
http://{\-}www.{\-}isis.{\-}rl.{\-}ac.{\-}uk/{\-}ess/{\-}neut\%5Fess.{\-}htm.

\bibitem{PDG98}
C. Caso \textit{et al.} (Particle Data Group),
Eur. Phys. J. \textbf{C3}, 1 (1998).

\bibitem{MSW}
S.P. Mikheyev and A.Yu. Smirnov,
Yad. Fiz. \textbf{42}, 1441 (1985)
[Sov. J. Nucl. Phys. \textbf{42}, 913 (1985)];
Il Nuovo Cimento \textbf{C9}, 17 (1986);
L. Wolfenstein,
Phys. Rev. \textbf{D17}, 2369 (1978);
Phys. Rev. \textbf{D20}, 2634 (1979).

\bibitem{Bugey-95}
B. Achkar \textit{et al.} (Bugey Coll.),
Nucl. Phys. \textbf{B434}, 503 (1995).

\bibitem{four-models}
J.T. Peltoniemi \textit{et al.}, 
Phys. Lett. \textbf{B298}, 383 (1993);
E.J. Chun \textit{et al.},
Phys. Lett. \textbf{B357}, 608 (1995);
S.C. Gibbons \textit{et al.},
Phys. Lett. \textbf{B430}, 296 (1998);
B. Brahmachari and R.N. Mohapatra,
Phys. Lett. \textbf{B437}, 100 (1998);
S. Mohanty \textit{et al.}, 
Phys. Lett. \textbf{B445}, 185 (1998);
J.T. Peltoniemi and J.W.F. Valle,
Nucl. Phys. \textbf{B406}, 409 (1993);
Q.Y. Liu and A.Yu. Smirnov,
Nucl. Phys. \textbf{B524}, 505 (1998);
D.O. Caldwell and R.N. Mohapatra,
Phys. Rev. \textbf{D48}, 3259 (1993);
E. Ma and P. Roy,
Phys. Rev. \textbf{D52}, R4780 (1995);
A.Yu. Smirnov and M. Tanimoto,
Phys. Rev. \textbf{D55}, 1665 (1997);
N. Gaur \textit{et al.},
Phys. Rev. \textbf{D58}, 071301 (1998);
E.J. Chun \textit{et al.}, 
Phys. Rev. \textbf{D58}, 093003 (1998);
K. Benakli and A.Yu. Smirnov,
Phys. Rev. Lett. \textbf{79}, 4314 (1997);
Y. Chikira, N. Haba and Y. Mimura,
hep-ph/9808254;
C. Liu and J. Song,
Phys. Rev. \textbf{D60}, 036002 (1999); 
W. Grimus, R. Pfeiffer and T. Schwetz,
hep-ph/9905320.

\bibitem{four-phenomenology}
J.J. Gomez-Cadenas and M.C. Gonzalez-Garcia,
Z. Phys. \textbf{C71}, 443 (1996); 
S. Goswami,
Phys. Rev. \textbf{D55}, 2931 (1997); 
V. Barger, T.J. Weiler and K. Whisnant,
Phys. Lett. \textbf{B427}, 97 (1998); 
V. Barger \textit{et al.}, 
Phys. Rev. \textbf{D59}, 113010 (1999); 
C. Giunti,
hep-ph/9906456.

\bibitem{BGKP-96}
S.M. Bilenky \textit{et al.}, 
Phys. Rev. \textbf{D54}, 4432 (1996). 

\bibitem{BGG-AB-96}
S.M. Bilenky, C. Giunti and W. Grimus,
Eur. Phys. J. \textbf{C1}, 247 (1998), hep-ph/9607372;
Proc. of Neutrino '96, Helsinki, June 1996,
p.~174,
hep-ph/9609343.

\bibitem{BGG-bounds}
S.M. Bilenky, C. Giunti and W. Grimus,
Phys. Rev. \textbf{D57}, 1920 (1998). 

\bibitem{BGG-CP}
S.M. Bilenky, C. Giunti and W. Grimus,
Phys. Rev. \textbf{D58}, 033001 (1998). 

\bibitem{Okada-Yasuda-97}
N. Okada and O. Yasuda,
Int. J. Mod. Phys. A \textbf{12}, 3669 (1997). 

\bibitem{BGGS-BBN-98}
S.M. Bilenky \textit{et al.}, 
Astropart. Phys. \textbf{11}, 413 (1999). 

\bibitem{BGG-BBN-conf}
S.M. Bilenky \textit{et al.}, 
in
\textit{New Trends in Neutrino Physics},
Proc. of the Ringberg Euroconference
(Tergernsee,
May 1998), 
p.~117,
hep-ph/9807569;
in
\textit{New Era in Neutrino Physics},
Proc. of a Satellite Symposium after Neutrino '98
(Tokyo, 
June 1998), 
p.~179,
hep-ph/9809466.	  

\bibitem{Barger-variations-98}
V. Barger \textit{et al.}, 
Phys. Rev. \textbf{D58}, 093016 (1998). 

\bibitem{BGGS-AB-99}
S.M. Bilenky \textit{et al.}, 
Phys. Rev. \textbf{D60}, 073007 (1999). 

\bibitem{BGKM-bb-98}
S.M. Bilenky \textit{et al.}, 
Phys. Rev. \textbf{D57}, 6981 (1998). 

\bibitem{BG-bb-98-99}
S.M. Bilenky, C. Giunti and W. Grimus,
Nucl. Phys. B (Proc. Suppl.) \textbf{77}, 151 (1999), hep-ph/9809368;
S.M. Bilenky and C. Giunti,
hep-ph/9904328.

\bibitem{Giunti-99-bb}
C. Giunti,
hep-ph/9906275.

\bibitem{BGGKP-bb-99}
S.M. Bilenky \textit{et al.}, 
hep-ph/9907234
[Phys. Lett. B, in press].

\bibitem{DGKK-99}
D. Dooling, C. Giunti, K. Kang and C.W. Kim, hep-ph/9908513.

\bibitem{Fuller-99}
G.C. McLaughlin \textit{et al.}, 
Phys. Rev. \textbf{C59}, 2873 (1999); 
A.B. Balantekin and G.M. Fuller,
hep-ph/9908465.

\bibitem{CDHS-CCFR-84}
F. Dydak \textit{et al.} (CDHS Coll.),
Phys. Lett. \textbf{B134}, 281 (1984);
I.E. Stockdale \textit{et al.} (CCFR Coll.),
Phys. Rev. Lett. \textbf{52}, 1384 (1984).

\bibitem{concha}
M.C. Gonzalez-Garcia,
private communication.

\bibitem{BHP80-Kobzarev80-Doi81-Langacker87}
S.M. Bilenky, J.~Ho\v{s}ek and S.T. Petcov,
Phys. Lett. \textbf{B94}, 495 (1980);
I.Yu. Kobzarev \textit{et al.},
Sov. J. Nucl. Phys. \textbf{32}, 823 (1980);
M. Doi \textit{et al.},
Phys. Lett. \textbf{B102}, 323 (1981);
P. Langacker \textit{et al.},
Nucl. Phys. \textbf{B282}, 589 (1987).

\bibitem{BBN-Nnu}
S. Burles \textit{et al.}, 
Phys. Rev. Lett. \textbf{82}, 4176 (1999); 
E. Lisi, S. Sarkar and F.L. Villante,
Phys. Rev. \textbf{D59}, 123520 (1999); 
S. Sarkar,
astro-ph/9903183;
K.A. Olive, G. Steigman and T.P. Walker,
astro-ph/9905320.

\end{thebibliography}
\end{document}